\documentclass[a4paper,twoside,10pt]{letter}
\usepackage{graphicx,saj,multicol,subeqnarray}

\def\papertype{}

\def\udc{...}
\begin{document}
\baselineskip=3.1truemm
\columnsep=.5truecm
\newenvironment{lefteqnarray}{\arraycolsep=0pt\begin{eqnarray}}
{\end{eqnarray}\protect\aftergroup\ignorespaces}
\newenvironment{lefteqnarray*}{\arraycolsep=0pt\begin{eqnarray*}}
{\end{eqnarray*}\protect\aftergroup\ignorespaces}
\newenvironment{leftsubeqnarray}{\arraycolsep=0pt\begin{subeqnarray}}
{\end{subeqnarray}\protect\aftergroup\ignorespaces}
\newcommand{\D}{$^\circ$}
\newcommand{\HI}{H\,{\sc i}}
\newcommand{\NaI}{Na\,{\sc i}}
\newcommand{\HeI}{He\,{\sc i}}
\newcommand{\HeII}{He\,{\sc ii}}
\newcommand{\FeVII}{[Fe\,{\sc vii}]}
\newcommand{\FeVI}{[Fe\,{\sc vi}]}
\newcommand{\FeV}{[Fe\,{\sc v}]}
\newcommand{\FeI}{Fe\,{\sc i}}
\newcommand{\FeII}{[Fe\,{\sc ii}]}
\newcommand{\FeIII}{[Fe\,{\sc iii}]}
\newcommand{\HII}{H\,{\sc ii}}
\newcommand{\CaII}{Ca\,{\sc ii}}
\newcommand{\hii}{h\,{\sc ii}}
\newcommand{\SII}{[S\,{\sc ii}]}
\newcommand{\SVI}{[S\,{\sc vi}]}
\newcommand{\SIII}{[S\,{\sc iii}]}
\def\p0{\phantom{0}}
\newcommand{\tbsp}{\rule{0pt}{10pt}}
\def\arcmin{\hbox{$^{\prime}$}}
\def\arcsec{\hbox{$^{\prime\prime}$}}

\newcommand{\OI}{[O\,{\sc i}]}
\newcommand{\OII}{[O\,{\sc ii}]}
\newcommand{\OIII}{[O\,{\sc iii}]}
\newcommand{\NII}{[N\,{\sc ii}]}
\newcommand{\NeIII}{[Ne\,{\sc iii}]}
\newcommand{\NeIV}{[Ne\,{\sc iv}]}
\newcommand{\ArIII}{[Ar\,{\sc iii}]}



\markboth{\eightrm The Strongest 100 Point Radio Sources in the LMC at 1.4~GHz
}
{\eightrm J.L. Payne et al. }

{\ }

\publ

\type

{\ }


\title{The Strongest 100 Point Radio Sources in the LMC at 1.4~GHz}


\authors{J. L. Payne$^{1}$, L. A. Tauber$^{1}$, M. D. Filipovi\'c$^{2}$, E. J. Crawford$^{2}$ and A. Y. De Horta$^{2}$  }

\vskip3mm

\address{$^1$Centre for Astronomy, James Cook University\break Townsville QLD, 4811, Australia}

\address{$^2$University of Western Sydney, Locked Bag 1797\break Penrith South, DC, NSW 1797, Australia}


\dates{January 16, 2009}{TBA, 2009}


\summary{We present the 100 strongest 1.4~GHz point sources from 
 a new mosaic image in the direction of the Large Magellanic Cloud (LMC). 
 The observations making up the mosaic were made over a ten year period and were combined with 
 Parkes single dish data at 1.4~GHz to complete the image for short spacing. 
 An initial list of co-idenfications within 10\arcsec\ at 0.843, 4.8 and 8.6~GHz 
 consisted of 2682 sources. Elimination of extended objects and artifact noise allowed 
 the creation of a refined list containing 1988 point sources. Most of these are presumed 
 to be background objects seen through the LMC; a small portion may represent 
 compact \HII\ regions, young SNRs and radio planetary nebulae. For the 1988 point sources 
 we 
 find a preliminary average spectral index ($\alpha$) of --0.53 and present a 1.4~GHz image 
 showing source location in the direction of the LMC.  
  } 


\keywords{Radio Continuum: galaxies -- Magellanic Clouds -- Catalogs}

\begin{multicols}{2}
{


\section{1. INTRODUCTION}

The Large Magellanic Cloud (LMC) is a galaxy, that from our vantage point, appears as a nearly face-on spiral. It is an ideal location to study and develop our understanding of radio sources such as, supernova remnants (SNRs), \HII\ regions, planetary nebulae, and \mbox{X-ray} binaries and their interactions. More distant radio sources that include active galaxies can also be seen in the direction of the LMC. These sources, that are not in the LMC, must also be identified and distinguished from those within the LMC.

We present a description and analysis of our data in Section~2 and the results in Section~3. Concluding remarks with a brief discussion are presented in Section~4. 


\section{2. OBSERVATIONS AND ANALYSIS}
 \label{observations}
 
During October 1994 and February 1995, the LMC was surveyed in mosaic mode, centred at 1.4~GHz (bandwidth 128~MHz), using the Australia Telescope Compact Array (ATCA). This survey divided the LMC into 12 regions, each containing 112 pointing centres. Each pointing centre was observed approximately 115 times (ranging from 95 to 140) for 15~s. The data was reduced using the {\sc miriad} software suite (Sault \& Killeen~2006). Data from  the 64-m Parkes radio telescope of the same region at 1.401~GHz ($\lambda$=21~cm) was used to fill in the missing short spacing allowing large-scale structure to be resolved. The Parkes data was obtained in the 1990's with beamwidth of 16.6\arcmin\ and rms noise of 30~mJy~beam$^{-1}$. As described in Hughes et al.~(2007), the resulting combined 21-cm mosaic image of the LMC has a resolution of $\sim40$\arcsec\ and a sensitivity of \mbox{$\sim0.3$~mJy~beam$^{-1}$}. Hughes at al. (2007) developed a pioneering technique for the more efficient cleaning of the side-lobes which are mainly created by strong sources such as 30~Doradus. 

Similar ATCA/Parkes mosiac images have been created from observations at 4.8 and 8.6~GHz ($\lambda$=6 and 3~cm) during the period between 2001 and 2003 by Dickel et al.~(2005). Their 4.8~GHz image has a FWHM of 33\arcsec\ while the 8.6~GHz image has a FWHM of 20\arcsec. Our analysis of these images shows a rms noise of 0.36 and 0.55~mJy~beam$^{-1}$ at 4.8 and 8.6~GHz, respectively. The original data was also re-processed by A.~Hughes using the same technique as discussed in Hughes et al.~(2007).

In our analysis, we also make use of radio surveys at 843~MHz made with the Molonglo Observatory Synthesis Telescope (MOST). This survey of the Magellanic Clouds has a angular resolution of 45\arcsec\ and a sensitivity of 0.7~mJy~beam$^{-1}$. Since this image covers the largest portion of sky second only to the 1.4~GHz image, its area represents the boundary limits for our catalogue. For more details, see Turtle \& Amy~(1991).

The radio images above were each analyzed using the {\sc imsad} task, which is part of the {\sc miriad} software suite. We initially selected all 1.4~GHz point sources with gaussian peak fluxes greater than 5$\sigma$ (1.5~mJy~beam$^{-1}$). The resulting list from {\sc imsad} were compared to those obtained using the 843 MHz, 4.8 and 8.6~GHz images with their respective 3$\sigma$ cutoff values. A preliminary point source catalogue was obtained for those 1.4~GHz sources having at least one co-identification within 10\arcsec\ at another frequency. 
The extent of the catalogue limited by the MOST observation region is \mbox{RA~(J2000)=06$^{h}$11$^{m}$00$^{s}$} to \mbox{04$^{h}$29$^{m}$00$^{s}$} and \mbox{DEC(J2000)=--64$^{0}$59$^{'}$00$^{''}$}to \mbox{--72$^{0}$49$^{'}$30$^{''}$}. This initial list contained 2682 objects. 

Each source was numbered and labeled using Karma's {\sc kvis} tool, which allowed for individual inspection. Sources that were extended at 1.4~GHz were excluded as well as those sources that appeared to be noise artifacts. The search area included the central portion of the LMC, although most sources there were extended; not appearing to be consistent with being point sources.

Using these latest images of the LMC we have found co-identifications with 56 previously identified radio SNRs from Filipovi\'c et al.~(1998) and an additional 20 candidate SNRs based on location, radio intensity, size and morphology. More details about the LMC SNRs study will be presented in the follow-up papers.

\section{3. RESULTS}
\label{results}

Using the methods described above, we were able to create a refined list containing an initial 1988 sources. 1.4~GHz flux densities (based on peak flux) range from 1.5~mJy to 1.5~Jy. Using   available flux densities ($S_{\nu}$) at all frequencies ($\nu$), individual spectral indices ($S_{\nu} \propto \nu^{\alpha}$) for each source was determined. From these we determine an  average spectral index ($\alpha$) of $-0.53$ (SD$=0.93$). Further analysis will most likely narrow this range as this list is compared to other known sources within the LMC.  
 
In Fig. 1, we show a histogram of the spectral indices for the strongest 100 point sources in our catalog. For these sources we see a symmetric curve indicating that most have indices between --0.8 and --1.0.   These values   range between indices of 0.2 and --2.2 and compares 
reasonably  to 
a similar histogram of background objects in the direction of the SMC as shown in Figure 1 of Payne et al. (2004). 
  
Table~1 gives source position (J2000; given in R.A. and Dec.), flux densities at 1.4~GHz, 843~MHz, 4.8~GHz and 8.6~GHz (given in Jy) and corresponding radio spectral index for the strongest 100 objects in the field of the LMC at 1.4 GHz. We show their positions in Fig.~2 (crosses). As expected, these sources cover the transparent region of the LMC evenly.
 
The expected background integral source count at 1.4~GHz was obtained by interpolating polynominial fits (log {\it N} -- log {\it S}) given by Wall~(1994). These fits give the predicted number of background sources in the observed field.  For 1.4~GHz, given a 5$\sigma$ cut-off of 1.5~mJy and a sky coverage of $\sim32$ square degrees\footnote{This represents the transparent region of the  843~MHz 
map  in the direction of the LMC.}, we expect to find 1960 background sources. This compares well to the 1988 sources reported here, especially since further comparisons with catalogues at other wavelengths will most likely eliminate a few more sources.

\section{4. CONCLUDING REMARKS AND SUMMARY}
 \label{summary}

As the LMC is relatively transparent in radio, background radio 
galaxies and quasars are seen throughout except for the densest regions. Most likely, 
some of these objects may alternatively represent compact \HII\ regions, microquasars, small SNRs and radio planetary nebulae. Comparisons with catalogues including these objects will 
be necessary before the completion of the final catalogue. 

We present here the strongest 100 sources from a preliminary analysis of a 1.4~GHz ATCA/Parkes 
mosiac image created by Hughes et al.~(2007). The refined list contains 1988 point objects having an average spectral index of --0.53; this 
list will be the subject of a future more comprehensive  paper. 

\acknowledgements{We used the Karma / {\sc miriad} software package developed by the ATNF. The Australia Telescope Compact Array is part of the Australia Telescope, which is funded by the Commonwealth of Australia for operation as a National Facility managed by CSIRO. }


\references


Dickel, J. R., McIntyre, V. J., Gruendl, R. A., Milne, D.: 2005, \journal{AJ}, \vol{129}, 790.

Filipovi\'c, M. D., Haynes, R. F., White, G. L., Jones, P. A.: 1998, \journal{A\&AS}, \vol{130}, 421.


Hughes, A., Staveley-Smith, L., Kim, S., Wolleben, M., Filipovi\'c, M. D.: 2007, \journal{MNRAS}, \vol{382}, 543.

Payne, J. L., Filipovi\'c, M. D., Reid, W., Jones, P. A., Staveley-Smith, L., White, G. L.: 2004, \journal{MNRAS}, \vol{355}, 44.

Sault, B., Killeen, N.: 2006, {\sc miriad} Users Guide, Australia Telescope National Facility.

Turtle, A. J., Amy, S. W.: 1991, \journal{IAUS}, \vol{148}, 114.

Wall J. V.: 1994, \journal{AuJPh}, \vol{47}, 625.

\endreferences

}

\end{multicols}
\newpage

 \vskip-0.2cm
\centerline{\includegraphics[width=10cm,angle=-90]{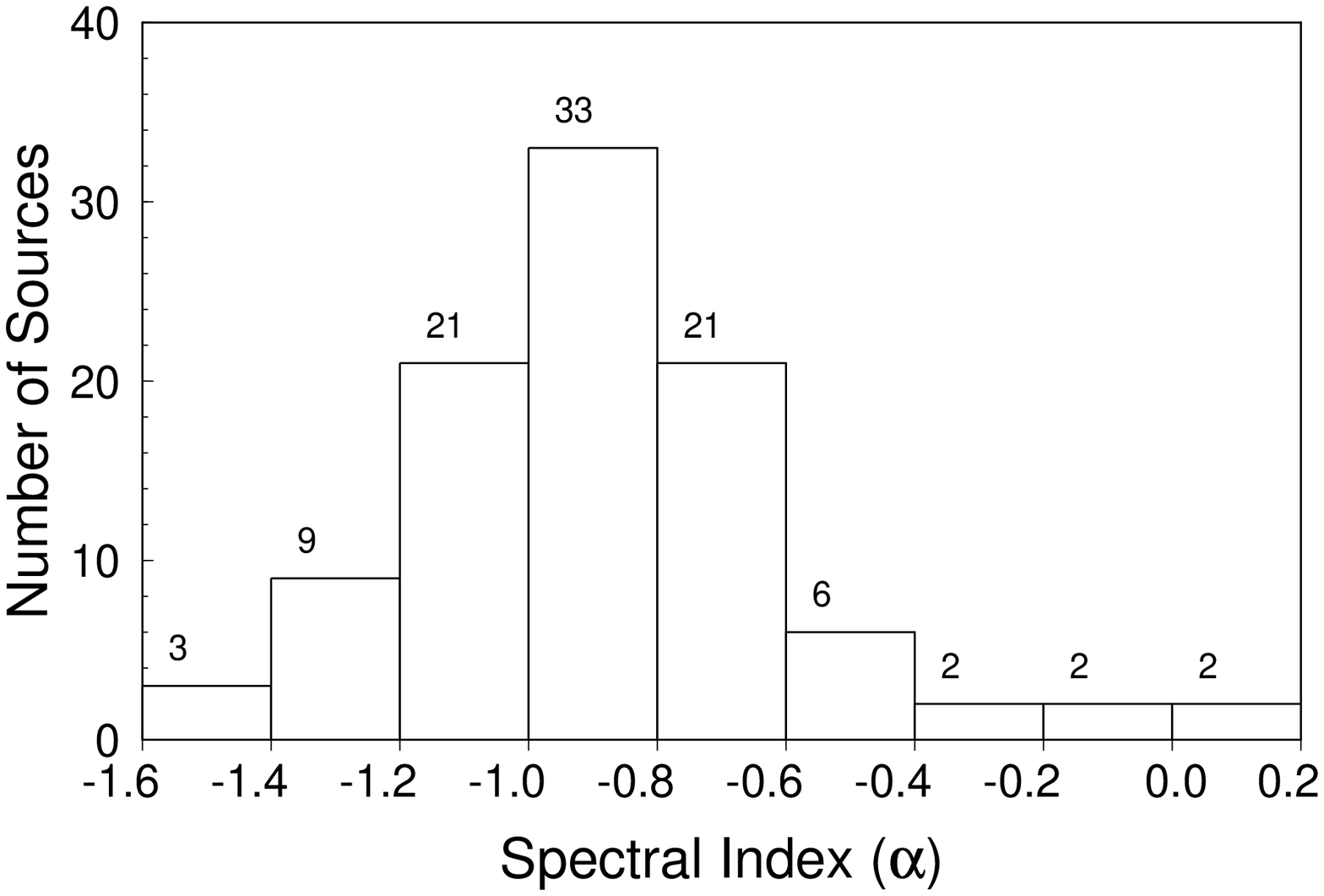}}
 \vskip-1cm
\figurecaption{1.}{ Histogram of spectral indices for the strongest 100 point radio sources in the field of the LMC. Only one object is outside of this graph range at $\alpha$=--2.2.}


\newpage
\vskip2.5cm

{{\bf Table 1.} Catalogue of the strongest 100 point sources in the field of the LMC at 1.4 GHz.}
\vskip2mm
\centerline{\begin{tabular}{|c|c|c|c|c|c|c|}
\hline
    R.A.        & Dec.      & 1.4 GHz & 843 MHz& 4.8 GHz&  8.6 GHz& Spectral Index \\
      (J2000.0)   & (J2000.0)     & (Jy)  &   (Jy)            &(Jy)& (Jy)& ($\alpha$)  \\
\hline
\noalign{\smallskip}
05:15:37.55	&	-67:21:28.1	&	1.477	&	2.27	&	0.5502	&	0.3548	&	--0.80	\\
06:00:05.11	&	-70:38:34.0	&	0.4854	&	0.9202	& -- & -- &	--1.26 	\\
05:45:54.05	&	-64:53:34.8	&	0.4593	&	0.6791	& -- & -- &	--0.77 	\\
05:28:10.63	&	-65:03:52.8	&	0.4487	&	0.6484	& -- & -- &	--0.73	\\
05:16:41.55	&	-71:49:05.3	&	0.3528	&	0.4766	&	0.1591	&	0.1144	&	--0.62 	\\
04:49:03.04	&	-70:52:04.6	&	0.3395	&	0.5204	&	0.1342	&	0.09757	&	--0.73	\\
05:32:54.25	&	-72:31:54.7	&	0.3013	&	0.4891	& -- & -- &	--0.96	\\
05:15:24.12	&	-65:58:41.1	&	0.2962	&	0.5108	&	0.07576	&	0.03643	&	--1.13 	\\
06:01:11.33	&	-70:36:08.3	&	0.2347	&	0.3005	& -- & -- &	--0.49	\\
05:29:30.11	&	-72:45:28.6	&	0.2294	&	0.262	& -- & -- &	--0.26	\\
05:16:37.78	&	-72:37:07.6	&	0.2224	&	0.323	& -- & -- &	--0.74	\\
05:52:05.93	&	-68:14:41.3	&	0.2219	&	0.3716	&	0.06859	&	0.04017	&	--0.96	\\
05:29:51.49	&	-67:49:33.2	&	0.2121	&	0.3251	& -- &	0.04381	&	--0.86	\\
05:44:27.14	&	-71:55:26.7	&	0.209	&	0.3359	&	0.05635	&	0.02649	&	--1.10	\\
04:46:11.45	&	-72:05:11.7	&	0.1865	&	0.2755	&	--	& -- &	--0.77	\\
06:00:02.78	&	-71:42:06.2	&	0.1825	&	0.3193	& -- & -- &	--1.10	\\
06:01:35.23	&	-69:55:45.2	&	0.1788	&	0.2933	& -- & -- &	--0.98	\\
05:05:01.10	&	-66:45:18.2	&	0.1751	&	0.3338	&	0.04099	&	0.01991	&	--1.21	\\
05:43:15.13	&	-68:06:52.0	&	0.1686	&	0.3298	&	0.03978	&	0.02182	&	--1.17	\\
05:33:44.77	&	-72:16:24.4	&	0.1554	&	0.1701	& -- & -- &	--0.18	\\
05:52:34.69	&	-66:40:41.6	&	0.1456	&	0.2902	& -- & -- &	--1.40	\\
04:49:38.98	&	-65:05:02.0	&	0.1436	&	0.2409	& -- & -- &	--1.02	\\
05:12:13.41	&	-72:32:47.4	&	0.1414	&	0.4356	& -- & -- &	--2.20\\
05:55:16.04	&	-67:20:51.1	&	0.1388	&	0.2429	& -- & -- &	--1.10	\\
04:37:44.59	&	-72:27:50.0	&	0.1354	&	0.1933	& -- & -- &	--0.70	\\
05:22:29.45	&	-70:37:55.0	&	0.1336	&	0.2112	&	0.04991	&	0.03348	&	--0.79	\\
04:37:04.43	&	-71:48:19.6	&	0.1321	&	0.2365	& -- & -- &	--1.15	\\
04:59:40.36	&	-69:55:03.7	&	0.1268	&	0.2037	&	0.03978	&	0.01401	&	--1.11	\\
05:59:57.73	&	-71:20:36.0	&	0.1258	&	0.2581	& -- & -- &	--1.42	\\
04:36:55.13	&	-69:30:33.1	&	0.1256	&	0.1943	& -- & -- &	--0.86	\\
04:56:08.72	&	-70:14:33.8	&	0.1256	&	0.1903	&	0.1527	&	0.1294	&	--0.09	\\
04:58:45.37	&	-72:50:24.7	&	0.1229	&	0.157	& -- & -- &	--0.48	\\
05:44:38.33	&	-65:34:54.2	&	0.1201	&	0.218	& -- & -- &	--1.18	\\
04:44:51.16	&	-67:46:36.2	&	0.1198	&	0.1967	& -- & -- &	--0.98	\\
04:45:12.98	&	-65:47:07.9	&	0.1148	&	0.1737	& -- & -- &	--0.82	\\
05:56:32.25	&	-71:29:06.0	&	0.1109	&	0.1804	&	0.03382	&	0.02142	&	--0.93	\\
05:01:39.04	&	-66:25:25.4	&	0.1094	&	0.1045	&	0.04726	&	0.02491	&	--0.63	\\
05:18:40.68	&	-65:36:14.8	&	0.1087	&	0.2151	& -- & -- &	--1.35	\\
04:44:16.21	&	-68:42:11.8	&	0.1081	&	0.1571	& -- & -- &	--0.74	\\
05:51:30.41	&	-69:16:31.0	&	0.1063	&	0.1736	&	0.03496	&	0.02236	&	--0.89	\\
04:38:51.46	&	-68:22:58.5	&	0.1049	&	0.1641	& -- & -- &	--0.88	\\
05:05:51.39	&	-69:51:17.4	&	0.1012	&	0.1487	&	0.05908	&	0.04966	&	--0.46	\\
06:00:35.26	&	-70:12:18.4	&	0.09992	&	0.1434	& -- & -- &	--0.71	\\
05:08:31.35	&	-67:06:16.4	&	0.09891	&	0.1666	&	0.03379	&	0.02128	&	--0.88	\\
05:47:45.43	&	-67:45:06.2	&	0.0956	&	0.1548	&	0.05575	&	0.03996	&	--0.55	\\
05:05:36.10	&	-70:05:14.8	&	0.09407	&	0.1571	&	0.02918	&	0.01682	&	--0.96	\\
05:00:36.49	&	-72:06:23.5	&	0.09076	&	0.1406	&	0.03057	&	0.01716	&	--0.90	\\
06:04:36.08	&	-71:02:23.6	&	0.09023	&	0.1273	& -- & -- &	--0.68	\\
05:05:31.12	&	-65:55:15.7	&	0.08999	&	0.1409	&	0.03114	&	0.01756	&	--0.89	\\
05:04:02.36	&	-72:03:45.5	&	0.08952	&	0.132	&	0.03226	&	0.01587	&	--0.90	\\
 \noalign{\smallskip}
\hline
\end{tabular}}

\vskip4cm


\newpage
\vskip2.5cm

{{\bf Table 1.} Catalogue of the strongest 100 point sources at 1.4 GHz (Continued).}
\vskip2mm
\centerline{\begin{tabular}{|c|c|c|c|c|c|c|}
\hline
    R.A.        & Dec.      & 1.4 GHz & 843 MHz& 4.8 GHz&  8.6 GHz& Spectral Index \\
      (J2000.0)   & (J2000.0)     & (Jy)  &   (Jy)            &(Jy)& (Jy)&  ($\alpha$) \\
\hline
\noalign{\smallskip}
05:05:39.65	&	-71:07:40.2	&	0.08849	&	0.138	&	0.03396	&	0.01994	&	--0.82	\\
05:51:40.46	&	-68:43:09.5	&	0.08817	&	0.1457	&	0.02994	&	0.02247	&	--0.82	\\
06:01:41.28	&	-72:38:32.5	&	0.0864	&	0.1225	& -- & -- &	--0.69	\\
05:43:17.55	&	-66:26:55.6	&	0.08633	&	0.05501	&	0.09289	&	0.09518	&	0.19	\\
04:40:08.66	&	-71:57:18.2	&	0.08249	&	0.1083	& -- & -- &	--0.54	\\
04:53:37.89	&	-68:29:27.8	&	0.08134	&	0.1089	&	0.05749	&	0.05211	&	--0.31	\\
04:43:18.45	&	-66:52:04.9	&	0.07911	&	0.1038	& -- & -- &	--0.54	\\
05:21:27.37	&	-65:21:40.3	&	0.0778	&	0.126	& -- & -- &	--0.95	\\
04:43:15.50	&	-66:02:48.5	&	0.07735	&	0.1349	& -- & -- &	--1.10	\\
04:51:19.90	&	-72:21:07.4	&	0.07439	&	0.1242	& -- & -- &	--1.01	\\
05:41:32.78	&	-67:06:15.7	&	0.07283	&	0.1241	&	0.0225	&	0.01355	&	--0.95	\\
05:56:13.83	&	-69:23:31.7	&	0.07255	&	0.1179	& -- & -- &	--0.96	\\
05:07:10.63	&	-71:44:04.8	&	0.07184	&	0.1414	&	0.009675	& -- &	--1.56	\\
05:36:36.32	&	-67:07:35.0	&	0.07069	&	0.1199	&	0.02514	& -- &	--0.89	\\
05:54:22.23	&	-67:54:50.8	&	0.06796	&	0.1138	& -- & -- &	--1.02	\\
04:55:51.47	&	-69:02:10.4	&	0.06759	&	0.07879	& -- &	0.09049	&	0.09	\\
05:09:31.11	&	-67:31:16.0	&	0.06748	&	0.09819	&	0.02745	&	0.02201	&	--0.66	\\
04:58:46.33	&	-67:05:38.5	&	0.06663	&	0.1061	&	0.02556	&	0.01788	&	--0.77	\\
05:48:45.09	&	-65:53:49.6	&	0.06527	&	0.1097	&	0.01753	&	0.007036	&	--1.16	\\
05:31:24.31	&	-65:16:34.0	&	0.06496	&	0.1088	& -- & -- &	--1.02	\\
04:42:11.81	&	-68:10:03.9	&	0.0645	&	0.08777	& -- & -- &	--0.61	\\
05:46:44.39	&	-66:41:14.2	&	0.06351	&	0.1038	&	0.02143	&	0.01385	&	--0.87	\\
05:46:36.81	&	-72:05:54.8	&	0.06315	&	0.08177	&	0.03052	&	0.01494	&	--0.70	\\
05:32:53.02	&	-67:09:42.0	&	0.06214	&	0.099	&	0.02105	&	0.0115	&	--0.92	\\
05:07:35.95	&	-71:52:34.6	&	0.06178	&	0.1044	&	0.01883	&	0.01067	&	--0.98	\\
05:19:28.01	&	-65:27:03.6	&	0.06149	&	0.1033	& -- & -- &	--1.02	\\
05:49:01.77	&	-71:20:11.4	&	0.0607	&	0.09022	&	0.02295	&	0.01599	&	--0.75	\\
05:07:17.62	&	-66:56:46.8	&	0.06061	&	0.1001	&	0.0193	&	0.01044	&	--0.96	\\
04:53:33.06	&	-70:40:27.2	&	0.05927	&	0.1036	&	0.02467	&	0.01444	&	--0.82	\\
05:55:01.85	&	-69:51:46.5	&	0.05892	&	0.1053	&	0.01232	& -- &	--1.24	\\
05:52:08.74	&	-70:31:30.2	&	0.05886	&	0.1049	&	0.01428	&	0.007234	&	--1.15	\\
04:52:56.95	&	-65:28:13.7	&	0.05784	&	0.09686	& -- & -- &	--1.02	\\
05:42:41.91	&	-66:02:57.5	&	0.0578	&	0.08926	&	0.02161	&	0.01046	&	--0.90	\\
05:06:34.32	&	-67:56:43.7	&	0.05736	&	0.09587	&	0.01923	&	0.01353	&	--0.85	\\
05:55:11.20	&	-69:47:51.4	&	0.05732	&	0.09435	&	0.01808	&	0.009905	&	--0.96	\\
05:41:27.27	&	-67:39:52.5	&	0.05719	&	0.1076	&	0.01316	& -- &	--1.21	\\
05:02:27.66	&	-65:57:11.7	&	0.05633	&	0.09523	&	0.01758	&	0.007245	&	--1.07	\\
05:14:07.63	&	-70:40:58.1	&	0.05559	&	0.09152	&	0.0139	&	0.004549	&	--1.26	\\
05:21:27.62	&	-67:07:23.3	&	0.05531	&	0.08736	&	0.02026	&	0.01078	&	--0.88	\\
05:56:53.22	&	-68:22:28.4	&	0.0546	&	0.1028	& -- & -- &	--1.25	\\
05:20:22.04	&	-72:44:44.7	&	0.05448	&	0.09054	& -- & -- &	--1.00	\\
05:12:30.38	&	-68:28:05.8	&	0.05408	&	0.07479	&	0.02454	&	0.01354	&	--0.72	\\
05:10:27.20	&	-69:32:09.1	&	0.05404	&	0.08677	&	0.01762	&	0.01143	&	--0.88	\\
05:18:13.26	&	-71:38:07.3	&	0.05333	&	0.07372	&	0.02262	&	0.01212	&	--0.76	\\
05:46:59.84	&	-65:35:00.9	&	0.05322	&	0.09416	& -- & -- &	--1.12	\\
05:51:05.04	&	-71:06:10.2	&	0.05313	&	0.09359	&	0.01655	&	0.01005	&	--0.96	\\
05:10:57.94	&	-65:29:34.2	&	0.05311	&	0.0727	& -- & -- &	--0.62	\\
06:03:47.39	&	-69:29:43.2	&	0.05301	&	0.08889	& -- & -- &	--1.02	\\ 
04:58:37.24      &       -67:12:44.2       &      0.0523       &      0.9973      &0.01316&0.003353&--1.39\\
05:04:09.69      & -67:01:10.8 & 0.05176&0.1005&0.0117&0.00481&--1.29\\
\noalign{\smallskip}
\hline
\end{tabular}}

\vskip4cm


\newpage
\centerline{\includegraphics[height=20cm,angle=270]{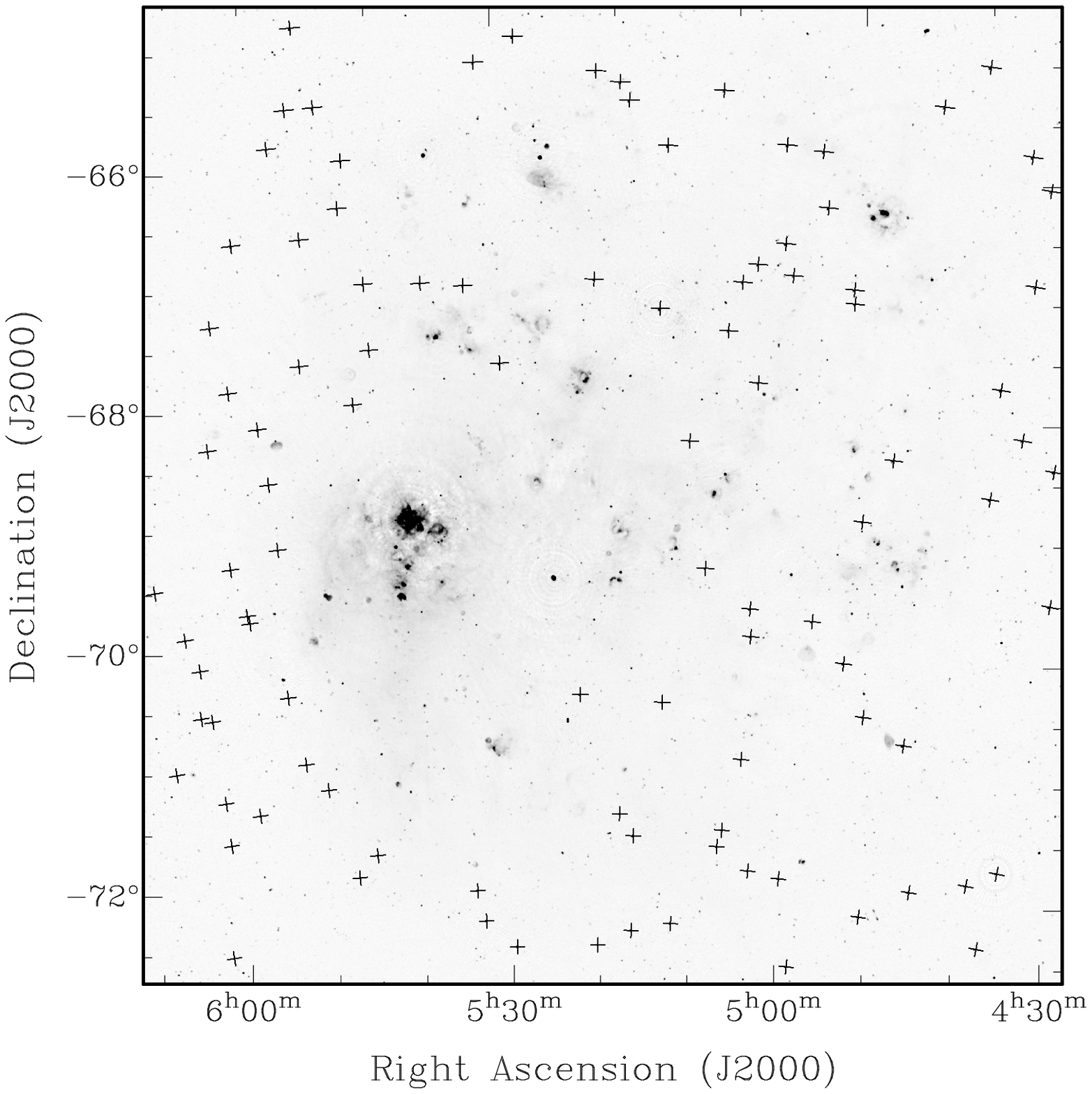}}

 
\figurecaption{2.}{Strongest 100 radio point sources (crosses) overlying the 1.4~GHz radio image of the LMC.  This image has 
a sensitivity of $\sim0.3$ mJy beam$^{-1}$ and a resolution of 40~\arcsec. }

\vfill\eject

{\ }



\naslov{100 NAJSNA{\ZZ}NIJIH TAQKASTIH RADIO OBJEKATA U VELIKOM MAGELANOVOM OBLAKU}


\authors{J. L. Payne$^{1}$, L. A. Tauber$^{1}$, M. D. Filipovi\'c$^{2}$, E. J. Crawford$^{2}$ and A. Y. De Horta$^{2}$  }

\vskip3mm

\address{$^1$Centre for Astronomy, James Cook University\break Townsville QLD, 4811, Australia}

\address{$^2$University of Western Sydney, Locked Bag 1797\break Penrith South, DC, NSW 1797, Australia}

\vskip.7cm


\centerline{\rrm UDK \udc}

\vskip1mm

\centerline{\rit Originalni nauqni rad}

\vskip.7cm

\begin{multicols}{2}

{

\rrm

U ovoj studiji predstav{lj}amo 100 najsna{\zz}nijih radio taqkastih objekata sa naxe nove mozaik slike Velikog Magelanovog Oblaka (VMO) na frekvenciji od 1.4~{\rm GHz}. Posmatra{nj}a korix{\cc}ena u ovoj studiji sakup{lj}ena su u posled{nj}ih 10 godina a tako{dj}e su i komplementovana sa Parks 64-m podatcima. Inicijalna lista objekata u krugu od 10\arcsec\ na frekvencijama od 0.843, 4.8 i 8.6~{\rm GHz} sastoji se od 2682 radio izvora. Eliminacijom netaqkastih objekata i artifekata xuma dovelo je do refiniranog kataloga od 1988 radio taqkastih objekta. Ve{\cc}ina ovih objekata su najverovatnije pozadinski objekti kao xto su kvazari ili aktivna galaktiqka jezgra koja se vide kroz transparentne delove VMO. Veoma mali procenat ovih objekata representuje objekte u samom VMO kao xto su {\rm H\,II} regioni, mladi ostatci supernovih i radio planetarne magline. Naxli smo da je sred{nj}i spektralni indeks za ovih 100 najsna{\zz}nijih radio izvora \mbox{$\alpha=$--0.53}. Takodje prezentujemo i uniformnu distribuciju ovih 100 objekata u po{lj}u VMO na 1.4~{\rm GHz}.

}
\end{multicols}

\end{document}